\documentclass[preprint,showkeys,preprintnumbers,amsmath,amssymb]{revtex4}

\usepackage[dvips]{graphicx}

\usepackage{amsmath}
  \usepackage{paralist}
  \usepackage{graphics} %% add this and next lines if pictures should be in esp format
  \usepackage{epsfig} %For pictures: screened artwork should be set up with an 85 or 100 line screen
\usepackage{graphicx}  \usepackage{epstopdf}%This is to transfer .eps figure to .pdf figure; please compile your paper using PDFLeTex or PDFTeXify.
 \usepackage[colorlinks=true]{hyperref}
\hypersetup{urlcolor=blue, citecolor=red}

\newtheorem{theorem}{Theorem}[section]
\newtheorem{corollary}{Corollary}
\usepackage{amssymb, amsmath}
\usepackage{mathrsfs}			% required for \lap := \mathscr L

\newcommand{\pf} {{\it Proof.}}
\newcommand{\qed} {{Q.E.D.}}

\newcommand{\Adj} {{\Lambda}}
 \newcommand{\ve}[1]{{\bf #1}}
 \newcommand{\vve}[1]{{\bf #1}}
 \newcommand{\veg}[1]{{\boldsymbol {#1}}}	% Greek vector
 \newcommand{\vveg}[1]{{\boldsymbol {#1}}}	% Greek matrix

\newcommand{\matthree}[9]{\bracket{\begin{array}{ccc}
		#1	&#2	&#3	\\
		#4	&#5	&#6	\\
		#7	&#8	&#9
		\end{array}}}

\newcommand{\mat}[4]{\bracket{\begin{array}{cc}
		#1	&#2\\
		#3	&#4	   	
		\end{array}}}

\newcommand{\D}{\partial}

\newcommand{\Dt}[1]{\frac {\D #1} {\D t}}

\newcommand{\dt}[1]{\frac {d #1} {d t}}
\newcommand{\delt}{{\Delta t}}
\newcommand{\hot}{o(\delt)}
\newcommand{\delw}{{\Delta\ve w}}

\newcommand{\dms}{n}		% dimension of the dynamical system
\newcommand{\ntime}{K}		% number of time steps in a series

\newcommand{\DI}[1]{\frac {\D #1} {\D x_1}}	%I, II, III, IV, V, etc.
\newcommand{\DII}[1]{\frac {\D #1} {\D x_2}}
\newcommand{\DIII}[1]{\frac {\D #1} {\D x_3}}
\newcommand{\Dn}[1]{\frac {\D #1} {\D x_\dms}}

\newcommand{\Di}[1]{\frac {\D #1} {\D x_i}}
\newcommand{\Dj}[1]{\frac {\D #1} {\D x_j}}
\newcommand{\DiDj}[1]{\frac {\D^2 #1} {\D x_i \D x_j}}
\newcommand{\DIDI}[1]{\frac {\D^2 #1} {\D x_1^2}}

\newcommand{\DDI}{\frac {\D} {\D x_1}}

\newcommand{\DDIDI}{\frac {\D^2} {\D x_1^2}}

\newcommand{\excl}[1]{{\backslash \hspace{-0.3em} #1}}
\newcommand{\others}{{2..\dms}}

\newcommand{\bracket}[1]{\left[#1\right]}

\newcommand{\parenth}[1]{\left(#1\right)}

\DeclareSymbolFont{AMSb}{U}{msb}{m}{n}
\DeclareMathSymbol{\R}{\mathbin}{AMSb}{"52}

\begin{document}

\title{Measuring the importance of individual units in producing 
	the collective behavior of a complex network}
%--------------------------------------------------------------------
%Measuring the effect of individual nodes on the structural integrity 
%of complex networks
%--------------------------------------------------------------------

\author{X. San Liang}
\email{X.S. Liang, sanliang@courant.nyu.edu}
\affiliation{Nanjing University of Information Science and Technology,
Nanjing, 210044, China}

%The abstract of your paper (less than 200 words)
\begin{abstract}
A quantitative evaluation of the contribution of individual units in
producing the collective behavior of a complex network can allow us to
understand the potential damage to the structure integrity due to the
failure of local nodes. Given time series for the units, a natural way to
do this is to find the information flowing from the unit of concern to the
rest of the network. In this study, we show that this flow can be
rigorously derived in the setting of a continuous-time dynamical system.
With a linear assumption, a maximum likelihood estimator can be obtained,
allowing us to estimate it in an easy way. As expected, this ``cumulative
information flow'' does not equal to the sum of the information flows to other individual units, reflecting the collective phenomenon that a group is not the addition of the individual members. For the purpose of demonstration and validation, we have examined a network made of Stuart-Landau oscillators. Depending on the topology, the computed information flow may differ. In some situations, the most crucial nodes for the network are not the hubs; they may have low degrees, and, if depressed or attacked, will cause the failure of the entire network.
\end{abstract}


\maketitle

% Enter the first author's name and address:
%\centerline{\scshape X. San Liang$^*$}

%\centerline{\scshape First-name2 last-name2 and First-name3
%last-name3}
%\medskip
%{\footnotesize
% % please put the address of the second  and third author
% \centerline{ First line of the address of the second author}
%   \centerline{Other lines}
%   \centerline{Springfield, MO 65810, USA}
%}

\bigskip

% The name of the associate editor will be entered by an editorial staff
% "Communicated by the associate editor name" is not needed for special issue.
% \centerline{(Communicated by the associate editor name)}
% \centerline{(Special issue: Mathematics for signal processing in complex
%	networks and systems)}

%\tableofcontents

\section{Introduction}

Complex networks provide a framework for the studies of 
many social, biological, and engineering systems 
such as the internet, brains, power grids, financial
trading markets, food webs, gene regulatory networks, to name a few.
A network consists of nodes or vertexes standing for the individual units 
or organizations, and links or edges for the interactions among the
nodes. For a node, the number of links connected to other nodes is called
its degree. By degree distribution we can have 
homogeneous and heterogeneous networks. The former class has binomial 
or Poisson degree distributions, examples including random
graphs\cite{Renyi1960} and small-world networks\cite{Strogatz1998}, 
while the latter class is scale free, bearing probability 
distributions $P$ of degree $k$ 
following a power law $P(k)\sim~k^{-\gamma}$, with an exponent
$\gamma\sim2-3$.
Most social\cite{Stanley2001}, biological\cite{Albert2005}, 
and technological networks\cite{Pastor2004} have the scale-free property;
other topological properties include high clustering coefficient, 
community and hierarchical structures, and, 
for directed networks, reciprocity, triad significance profile, etc.

A goal of complex network studies is to understand how 
individuals collaborate to produce the collective behavior. 
One question to ask is whether the connectivity of a network is robust 
to local node failure, deterioration or functional depression. Of
particular interest is whether initially
a tiny shock may cascade to disrupt the network on a large scale.
How to quantify the contribution of a unit to the network as a whole
is thence an important issue; it is related to many real world problems 
such as power grid failure (e.g., the 2003 massive
blackout that darkened much of the North American upper Midwest and
Northeast\cite{Fairley2004}),
control of epidemic disease, 
identification of bottlenecks in city traffic, etc. 
Usually this is studied by observing the connectivity after
preferential removal of a unit, which is found to have different effects 
on the two types of networks.
If the removal or attack is random, heterogeneous networks are quite 
robust as compared to homogeneous networks; if, however, the attack is 
intentional at some special nodes, then heterogeneous networks could be
rather fragile. These special nodes are usually highly connected ones, 
i.e., hubs, as easily imagined. 
Recently, Tanaka et al.\cite{Tanaka2012} observed that, sparsely 
connected nodes may be more important which, if functionally depressed, 
may result in drastic change in network structure.
That is to say, the structure integrity or robustness could also 
be largely influenced by low-degree nodes, rather than by hubs.
We hence cannot judge the importance of a unit simply by degree. 
It depends on many different properties of the network topology in question.

As said above, the problem is usually tackled by removing a unit and 
observing the change in topology of the network of concern. 
However, in many networks, biological networks in particular, this is
often infeasibloften infeasible, as breaking a unit means terminating the experiment.
On the other hand, we may have time series of measurements.
So the whole problem is converted into assessing the importance of a unit
from analyzing the signals as observed.
Previously, we have rigorously formulated information flow within 
dynamical systems(e.g., \cite{Liang2014}\cite{Liang2016});
it has been widely used for studying the causal
relations among dynamical events, and hence is readily for the study of the
interactions among nodes in a network.
One may think that the contribution of a given node may be obtained by
adding up all the information flows from it to the other nodes. 
Unfortunately, as we will see soon in the following sections, 
this is true only when all the nodes are disconnected, 
i.e., when the nodes do not form a network and hence no collective
behavior emerge.  
This from one aspect manifests the well-known fact that 
groups are not simply the addition of their individual members; 
they could be more or less (some social science examples can be seen
in \cite{Aleta2019}\cite{Baumeister2016}\cite{Malone2010}\cite{Mason2010}).

In the following, we first present the setting for the problem, and then
derive the information flow from an individual unit to the network.
Maximum likelihood estimation is made in section~\ref{sect:mle}; it 
yields a formula for easy assessment of the importance of a
node from given time series. As a validation, and also a demonstration 
of application, section~\ref{sect:landau} presents a network of
synchronized Stuart-Landau oscillators which, when a fraction of nodes
become deteriorated, may become silent completely. This study is concluded
in section~\ref{sect:summary}.

\section{Information flow from a unit to the entire network}

Consider a network modeled by an $\dms$-dimensional dynamical system
	\begin{eqnarray}	\label{eq:gov}
	\dt {\ve x} = \ve F (\ve x, t) + \vve B(\ve x, t) \dot{\ve w},
	\end{eqnarray}
where $\ve x$ is the state variable vector for the $\dms$ nodes
$(x_1, x_2,..., x_\dms)$, $\ve x \in\R^\dms$,
$\ve F = (F_1, ..., F_n)$ the differentiable functions of $\ve x$ and time $t$
describe the interaction paths (edges/links),
$\ve w$ is a vector of $m$ independent standard Wienner processes,
and $\vve B = (b_{ij})$ an $\dms\times m$ is the matrix of stochastic 
perturbation amplitude. Here we follow the convention in physics not to
distinguish a random variable and a deterministic variable. (In probability
theory, they are usually distinguished by upper-case and low-case symbols.)
To examine the influence of a unit to the entire network made of the $n$
units, it suffices to consider the component $x_1$; if not, we can always
re-arrange the vector $\ve x$ to make it so.
The whole problem now boils down to finding the information flow from $x_1$
to $(x_2, x_3, ..., x_\dms)$, which we will be denoting as $\ve x_\others$
henceforth (i.e., as $\ve x$ with component $1$ removed).

In \cite{Liang2016}, the information flow between two individual components 
$x_i$ and $x_j$ has been rigorously derived from first principles. 
But the information flow from one component, here $x_1$, to a multitude of
components, here $\ve x_\others$, is yet to be implemented. 
One may conjecture that it is just an addition of all flows from $x_1$ to all the
individual components of $\ve x_\others$. As we will see soon below, 
this is generally not the case, and the nonadditivity is a reflection of
the macrostate or collective behavior of a multi-connected network.

We follow the strategy used in \cite{Liang2008} to do the derivation. 
The information flow is, by the physical argument therein, the amount of 
entropy transferred from $x_1$ to $\ve x_\others$. We hence need to find the
evolution of the joint entropy of $\ve x_\others$, and single out the
contribution to this evolution from $x_1$. This result follows.

	\begin{theorem}
	For the dynamical system (\ref{eq:gov}), if the probability 
	density function (pdf) of $\ve x$ is compactly supported, then
	the information flow from $x_1$ to $(x_2,x_3,...,x_\dms)$ is
	\begin{eqnarray}	\label{eq:T1n}
	T_{1\to\others} 
	= -E\bracket{\sum_{i=2}^\dms \frac1 {\rho_\others}
				     \Di{F_i\rho_\others}  }
	  + \frac12 E\bracket{
		\sum_{i=2}^\dms \sum_{j=2}^\dms \frac 1 {\rho_\others}
			\DiDj {g_{ij} \rho_\others }
			     }.
	\end{eqnarray}
	The units are nats per unit time. 
	In the equation, $\rho_\others$ is joint pdf of $(x_2,x_3,...,x_\dms)$,
	$g_{ij} = \sum_{k=1}^m b_{ik} b_{jk}$,
	and $E$ signifies mathematical expectation.
	\end{theorem}

%\begin{proof} 	% proof for (eq:T1n)
\pf
Associated with (\ref{eq:gov}) there is a Fokker-Planck equation governing
the evolution of the pdf $\rho$ of $\ve x$:
	\begin{eqnarray}	\label{eq:fk}
	\Dt\rho + \DI {\rho F_1} + \DII {\rho F_2} + ... + \Dn{\rho F_n}
	= \frac 12 \sum_{i=1}^d \sum_{j=1}^n \DiDj {g_{ij}\rho},
	\end{eqnarray}
where $g_{ij} = \sum_{k=1}^m b_{ik} b_{jk}$, $i,j=1,...,\dms$. 
This marginal pdf of $x_1$, $\rho_1(x_1)$, 
is obtained by integrating out $(x_2,...,x_\dms)$ in 
(\ref{eq:fk}). By the assumption of compactness of $\rho$, the resulting
equation becomes
	\begin{eqnarray}
	\Dt {\rho_1} + \DDI \int_{\R^{\dms-1}} \rho F_1 d\ve x_\others
	= \frac12 \DDIDI \int_{\R^{\dms-1}} g_{11}\rho d\ve x_\others.
	\end{eqnarray}
For the sake of notational simplicity, here
we have written $dx_2 dx_3 ... dx_\dms$ as $d\ve x_\others$.
From this the evolution of the marginal entropy of $x_1$, written $H_1$,
can be derived:
	\begin{eqnarray}
	\dt {H_1} = -E\bracket{F_1 \DI {\log\rho_1} }
		    -\frac12 E\bracket{g_{11} \DII {\log\rho_1} }.
	\end{eqnarray}
See Liang (2008) for a proof.

To study impact of $x_1$ on the rest of the network, we need to consider
the evolution of the joint entropy of $(x_2,x_3,...x_\dms) = \ve x_\others$,
i.e., 
	$$H_\others= - \int_{\R^{\dms-1}} \rho_\others \log \rho_\others 
	d\ve x_\others,$$ 
where $\rho_\others = \rho_\others(x_2,...,x_\dms)
= \int_\R \rho dx_1$ is the joint pdf of $(x_2,x_3,...x_\dms)$.
By integrating out $x_1$ from Eq.~\ref{eq:fk}, we have 
	\begin{eqnarray}
	\Dt {\rho_\others} + \DII\ \int_\R \rho F_1 dx_1 + \hdots
		+ \Dn\ \int_\R \rho F_\dms dx_1
	= \frac12 \sum_{i=2}^\dms \sum_{j=2}^\dms 
		  \DiDj\ \int_\R g_{ij}\rho dx_1.
	\end{eqnarray}
Multiply $-(1+\log\rho_\others)$, then integrate over $\R^{\dms-1}$.
The first term is $dH_\others/dt$. By taking advantage of the compactness
assumption, the second term on the left hand side results in
	\begin{eqnarray*}
	&&- \int_{\R^{\dms-1}} \bracket{(1+\log\rho_\others) 
		\DII\ \parenth{\int_\R\rho F_2 dx_1}  } d\ve x_\others	\\
	&&= -\int_{\R^{\dms-1}} \log\rho_\others \DII\ 
		\parenth{\int_\R \rho F_2 dx_1} d\ve x_\others	\\
	&&= \int_{\R^{\dms-2}} \left\{
\left[-\log\rho_\others \cdot \int_\R \rho F_2 dx_1\right]_{-\infty}^\infty
+ \int_\R \parenth{\int_\R \rho F_2 dx_1} \cdot \DII{\log\rho_\others} dx_2
				\right\} dx_3...dx_\dms  	\\
%	&&\flushright {\rm (integration\ by\ parts)}	\\
	&&= \int_{\R^\dms} \rho F_2 \DII {\log\rho_\others} d\ve x 
	  = E\bracket{F_2 \DII {\log\rho_\others} },
	\end{eqnarray*} 
where $E$ signifies mathematical expectation.
Likewise, the third term through the $\dms^{th}$ term are
	\begin{eqnarray*}
	 E\bracket{F_3 \DIII {\log\rho_\others} }, \
	\hdots, \
	 E\bracket{F_\dms \Dn {\log\rho_\others} }.
	\end{eqnarray*}
On the right hand side, the $(i,j)^{\rm th}$ component is
	\begin{eqnarray*}
	&&-\int_{\R^{\dms-1}} \bracket{  
	  (1+\log\rho_\others) \cdot \frac12 \DiDj\ \int_\R g_{ij}\rho dx_1
				 }d\ve x_\others		\\
	&&=-\frac12 \int_{\R^{\dms-1}} \log\rho_\others \cdot
	  \DiDj\ \parenth{\int_\R g_{ij} \rho dx_1}  d\ve x_\others  \\
	&&=-\frac12 \int_{\R^{\dms-2}}
	   \left\{
	      \bracket{\log\rho_\others \cdot \Dj\ 
	         (\int_\R g_{ij}\rho dx_1) }_{-\infty}^\infty  \right.\\
	&&\qquad\qquad\qquad
	         \left. - \int_\R \Di {\log\rho_\others} \cdot 
			\Dj {\int_\R g_{ij}\rho dx_1}
	      dx_i	
	   \right\}  dx_2...dx_{i-1}dx_{i+1}...dx_\dms		\\
	&&=\frac12 \int_{\R^{\dms-1}} \Di {\log\rho_\others} 
		    \cdot \Dj {\int_\R g_{ij}\rho dx_1}  d\ve x_\others \\
	&&=\frac12 \int_{\R^{\dms-2}}
	   \left\{
       	        \bracket{\Di {\log\rho_\others} \int_\R g_{ij} 
				\rho dx_1}_{-\infty}^\infty   \right.	\\
	&&\qquad\qquad\qquad
		\left.
		- \int_\R \parenth{\int_R g_{ij}\rho dx_1} \cdot
			\DiDj {\log\rho_\others} dx_j
	   \right\} dx_2...dx_{j-1}dx_{j+1}...dx_\dms		\\
	&&
	  =-\frac12 \int_{\R^\dms} \rho g_{ij} \DiDj {\log\rho_\others} d\ve x
	  =-\frac12 E\bracket{g_{ij} \DiDj {\log\rho_\others} }.
	\end{eqnarray*}
Putting the above together, we have
	\begin{eqnarray}	\label{eq:Hother}
	\dt {H_\others} = -\sum_{i=2}^\dms E \bracket{F_i \Di{\log\rho_\others}}
			- \frac12 \sum_{i=2}^\dms \sum_{j=2}^\dms
			  E\bracket{g_{ij} \DiDj{\log\rho_\others} }.
	\end{eqnarray}

The evolution of $H_\others$ contains two parts, 
one being the effect of $x_1$, another being the part with the effect of
$x_1$ excluded. We denote the latter by $dH_{\others,\excl1}/dt$;
it can be found by instantaneously freezing $x_1$ as a parameter.
For this purpose, we examine, on an infinitesimal interval
$[t,~t+\delt]$, a system modified from the original (\ref{eq:gov}) 
by removing its first equation, i.e.,
	\begin{eqnarray}
	&&\dt{x_2} = F_2(x_1,x_2,...,x_\dms; t) 
		+ \sum_{k=1}^m b_{2k}(x_1,x_2,...,x_\dms; t) \dot w_k\\
	&&\dt{x_3} = F_3(x_1,x_2,...,x_\dms; t) 
		+ \sum_{k=1}^m b_{3k}(x_1,x_2,...,x_\dms; t) \dot w_k\\
	&&\quad\vdots	\qquad\qquad\qquad\qquad\qquad\quad\ \ \vdots \cr
	&&\dt{x_\dms} = F_\dms(x_1,x_2,...,x_\dms; t) 
		+ \sum_{k=1}^m b_{\dms k}(x_1,x_2,...,x_\dms; t) \dot w_k.
	\end{eqnarray}
Note here the $F_i$'s and $b_{ik}$'s still have dependence on $x_1$, but
now $x_1$ appears in the modified system as a parameter. Given the pdf 
of $\ve x$ at time $t$, we need to find the pdf of $\ve x_\excl1$ at 
time $t+\delt$.
In Liang (2016), this is fulfilled by first constructing a
mapping $\Phi: \R^{\dms-1} \to \R^{\dms-1}$, $\ve x_\excl1(t) \mapsto \ve
x_\excl1(t+\delt)$, then studying the Frobenius-Perron
operator of the modified system.
Here we choose an alternative approach.
Note on the interval $[t, t+\delt]$, there also exists a Fokker-Planck
equation for the modified system
	\begin{eqnarray}	\label{eq:fk_modified}
	&&
	\Dt {\rho_\excl1} + \DII {F_2\rho_\excl1} + \DIII {F_3\rho_\excl1}
			  + \hdots + \Dn {F_\dms\rho_\excl1}
	= \frac12 \sum_{i=2}^\dms \sum_{j=2}^\dms \DiDj {g_{ij}\rho_\excl1},\\
	&&
	\rho_\excl1 = \rho_\others \qquad\qquad\qquad {\rm at\ time\ t.}
	\end{eqnarray}
Here $g_{ij} = \sum_{k=1}^m b_{ik} b_{jk}$ is still as before;
$\rho_\excl1$ means the joint pdf of $(x_2,...,x_\dms)$ with $x_1$
frozen as a parameter. 
$\rho_\excl1$ is somehow similar to the conditional pdf of the
former on the latter, but not exactly as that. The subscript $\excl1$
signifies that $x_1$ is removed from the independent variables. Note this
is quite different from $\rho_\others$, which has no dependence on $x_1$ at
all; but they are equal at time $t$.

Divide (\ref{eq:fk_modified}) by $\rho_\excl1$ to get
	\begin{eqnarray*}
	\Dt {\log\rho_\excl1} + 
	   \sum_{i=2}^\dms \frac 1{\rho_\excl1} \Di{F_i\rho_\excl1} 
	= \frac1{2\rho_\excl1} 
	  \sum_{i=2}^\dms\sum_{j=2}^\dms \DiDj {g_{ij}\rho_\excl1}.
	\end{eqnarray*}
Discretizing, and noticing that $\rho_\excl(t) = \rho_\others(t)$,
we have
	\begin{eqnarray*}
	&& \log\rho_\excl(\ve x_\excl1; t+\delt)	\\
	&& 
	= \log\rho_\others(\ve x_\excl1; t) 
	  - \delt \cdot \sum_2^\dms \frac 1{\rho_\others} \Di {F_i\rho_\others}
	  + \frac\delt 2 \sum_2^\dms \sum_2^\dms \frac 1 {\rho_\others}
			\DiDj {g_{ij}\rho_\others}
	  + \hot.
	\end{eqnarray*}
To arrive $dH_{\others,\excl1}/dt$, we need to find 
	$\log\rho_\excl(\ve x_\excl1(t+\delt); t+\delt)$. 
Using the Euler-Bernstein approximation,
	\begin{eqnarray}
	\ve x_\excl1(t+\delt) = \ve x_\excl1(t) + \ve F_\excl1\delt 
		+ \vve B_\excl1 \delw,
	\end{eqnarray}
where, just like the notation $\ve x_\excl1$, 
	\begin{eqnarray*}
	&& \ve F_\excl1 = (F_2,...,F_\dms)^T,	\\
	&& \vve B_\excl1 = \matthree
			   {b_{21}} \hdots  {b_{2m}}
			   \vdots   \ddots  \vdots
			   {b_{\dms1}} \hdots  {b_{\dms m}} \\
	&& \delw = (\Delta w_1, ..., \Delta w_m)^T
	\end{eqnarray*}
and $\Delta w_k \sim N(0, \delt)$,
we have 
	\begin{eqnarray*}
	&&\log(\rho_\excl1(\ve x_\excl(t+\delt); t+\delt)	\\
	&&
	= \log\rho_\others(\ve x_\excl1(t) + \ve F_\excl1\delt 
					  + \vve B_\excl1\delw; t)  \\
	&&\ \ \
	  - \delt \cdot \sum_2^\dms \frac 1{\rho_\others} \Di {F_i\rho_\others}
	  + \frac\delt 2 \sum_2^\dms \sum_2^\dms \frac 1 {\rho_\others}
			\DiDj {g_{ij}\rho_\others}
	  + \hot.	\\
	&&
	= \log\rho_\others(\ve x_\excl1(t)) 
	  + \sum_{i=2}^\dms \bracket{
	    \Di{\log\rho_\others} (F_i\delt + \sum_{k=1}^m b_{ik} \Delta w_k)
				   }		\\
	&&\ \ \
	  + \frac12\cdot \sum_{i=2}^\dms \sum_{j=2}^\dms
	    \bracket{
            \DiDj {\log\rho_\others} (F_i\delt + \sum_{k=1}^m b_{ik}\Delta w_k)
                             \cdot  (F_j\delt + \sum_{l=1}^m b_{jl}\Delta w_l)
		    }		\\
	&&\ \ \
	  - \delt \cdot \sum_2^\dms \frac 1{\rho_\others} \Di {F_i\rho_\others}
	  + \frac\delt 2 \sum_2^\dms \sum_2^\dms \frac 1 {\rho_\others}
			\DiDj {g_{ij}\rho_\others}
	  + \hot.
	\end{eqnarray*}
Take mathematical expectation on both sides. The left hand side is
$-H_{\others, \excl1}(t+\delt)$.
By the Corollary III.I of Liang (2016), and noting 
$E\Delta w_k = 0$, $E\Delta w_k^2 = \delt$ and the fact that 
$\delw$ are independent of $\ve x_\excl1$, we have
	\begin{eqnarray*}
	&& - H_{\others, \excl1}(t+\delt) = -H_\others(t) + 
	\delt \cdot E\sum_{i=2}^\dms F_i \Di{\log\rho_\others}\\
	&&\qquad 
	  + \frac\delt 2 \cdot E \sum_{i=2}^\dms \sum_{j=2}^\dms
	 	\sum_{k=1}^m \sum_{l=1}^m b_{ik} b_{jl} \delta_{kl}
			\DiDj {\log\rho_\others}	\\
	&&\qquad
	  - \delt \cdot E \sum_2^\dms \frac 1{\rho_\others} \Di {F_i\rho_\others}
	  + \frac\delt 2 E \sum_2^\dms \sum_2^\dms \frac 1 {\rho_\others}
			\DiDj {g_{ij}\rho_\others} + \hot	\\
	&&=
	-H_\others(t) + \delt \cdot E\sum_{i=2}^\dms F_i \Di{\log\rho_\others}
	  + \frac\delt 2 \cdot E \sum_{i=2}^\dms \sum_{j=2}^\dms
	 	g_{ij} \DiDj {\log\rho_\others}	\\
	&&\qquad 
	  - \delt \cdot E \sum_2^\dms \frac 1{\rho_\others} \Di {F_i\rho_\others}
	  + \frac\delt 2 E \sum_2^\dms \sum_2^\dms \frac 1 {\rho_\others}
			\DiDj {g_{ij}\rho_\others} + \hot.
	\end{eqnarray*}
So
	\begin{eqnarray*}
	&& \dt {H_{\others,\excl1}} 
	= \lim_{\delt\to0} \frac {H_{\others,\excl1} - H_\others(t)} \delt\\
	&&\qquad
	= -E \sum_{i=2}^\dms \parenth{F_i \Di{\log\rho_\others} - 
			     \frac1{\rho_\others} \Di{F_i\rho_\others}}  \\
	&&\qquad\quad
	  -\frac12 E \sum_{i=2}^\dms \sum_{j=2}^\dms
		\parenth{g_{ij} \DiDj {\log\rho_\others} + 
			\frac1 {\rho_\others} \DiDj {g_{ij}\rho_\others}  }.
	\end{eqnarray*}
Hence the information flow from $x_1$ to $\ve x_\excl1$ is
	\begin{eqnarray*}% 	\label{eq:T1n}
	&&T_{1\to\others} = \dt {H_\others} - \dt {H_{\others,\excl1}}	\cr
	&&\qquad
	=-E \sum_{i=2}^\dms \parenth{F_i \Di {\log\rho_\others}  }
	 -\frac12 E \sum_{i=2}^\dms \sum_{j=2}^\dms 
		\parenth{g_{ij} \DiDj {\log\rho_\others}  }	\cr
	&&\qquad\ \ \ \
	 -E \sum_{i=2}^\dms \Di {F_i} 
	 + \frac12 E \sum_{i=2}^\dms \sum_{j=2}^\dms 
	   \parenth{g_{ij}\DiDj {\log\rho_\others} 
		  + \frac1{\rho_\others} \DiDj {g_{ij}\rho_\others} } \cr
	&&\qquad
	= -E\bracket{\sum_{i=2}^\dms \frac1 {\rho_\others}
				     \Di{F_i\rho_\others}  }
	  + \frac12 E\bracket{
		\sum_{i=2}^\dms \sum_{j=2}^\dms \frac 1 {\rho_\others}
			\DiDj {g_{ij} \rho_\others }
			     }.
	\end{eqnarray*}
%\end{proof}	% proof for (eq:T1n)
 \qed

There is a nice property regarding noise:
when the noise is additive, the stochastic contribution to the information
flow vanishes, as stated in the following corollary. 
	\begin{corollary} \label{cor:additive_noise}
	In (\ref{eq:gov}), if $\vve B$ does not depend on $\ve x$, then
	\begin{eqnarray*}
	T_{1\to\others}
	= -E\bracket{\sum_{i=2}^\dms \frac1 {\rho_\others}
				     \Di{F_i\rho_\others}}.
	\end{eqnarray*}
	\end{corollary}

%\begin{proof}
\pf
If $b_{ij}$ is independent of $\ve x$, so is 
$g_{ij} = \sum_{k=1}^m b_{ik}b_{jk}$. Thus,
	\begin{eqnarray*}
	&& E\sum_i\sum_j \frac 1 {\rho_\others} \DiDj {g_{ij} \rho_\others}
	   = \sum_i\sum_j g_{ij} \int_{R^\dms} \DiDj {\rho_\others} d\ve x \\
	&& = \sum_i\sum_j g_{ij} \int_{\R^{\dms-1}} 
		\frac {\int_\R \rho dx_1} {\rho_\others} \DiDj {\rho_\others}
		dx_2 dx_3...dx_\dms	\\
	&& = \sum_i\sum_j g_{ij} \int_{\R^{\dms-1}} \DiDj{\rho_\others}
		dx_2 dx_3...dx_\dms,
	\end{eqnarray*} 
which is zero by the compactness of $\rho$.
%\end{proof}
\qed

The formula (\ref{eq:T1n}) can be verified with the particular situation in
which the rest of the network does not depend on $x_1$. In this case $x_1$
plays no role. Indeed, if we follow the procedure for the above
corollary, it is easy to prove that $T_{1\to\others}$ vanishes. So we have:
	\begin{theorem} ({\bf Principle of nil causality})\ 
	If $\ve F_\excl1$ and $\vve B_\excl1$
	are independent of $x_1$, $T_{1\to\others} = 0$.
	\end{theorem}

\subsection{Linear systems}

Steered by a linear system, a Gaussian process is always Gaussian.
In this case, the information flow can be greatly simplified.

	\begin{theorem}
	In (\ref{eq:gov}), suppose
	\begin{eqnarray}
	&& F_i = f_i + \sum_{j=1}^\dms a_{ij} x_j,
	\end{eqnarray}
	where $f_i$ and $a_{ij}$ are constants, and $b_{ij}$ are also constants.
	Further suppose that initially $\ve x$ has a Gaussian distribution, 
	then
	\begin{eqnarray}	\label{eq:T1n_linear}
	T_{1\to\others} = \sum_{i=2}^\dms
	\bracket {
	\sum_{j=2}^\dms \sigma_{ij}' 
		\parenth{\sum_{k=1}^\dms a_{ik} \sigma_{kj}}
		- a_{ii},
		 }
	\end{eqnarray}
	where $\sigma_{ij}'$ is the $(i,j)^{\rm th}$ entry of
	$\mat 1 {\vve 0} {\vve 0} {\vveg\Sigma_\excl1^{-1}}$.
	\end{theorem}

%\begin{proof}
\pf
In (\ref{eq:T1n}),
by Corollary~\ref{cor:additive_noise}, the stochastic part 
(second term) can be ignored.
Suppose the joint pdf of $\ve x$ has a form like
	\begin{eqnarray}
	\rho(x_1,...,x_\dms) = \frac 1 {\sqrt{(2\pi)^\dms \det\vveg\Sigma}}
	e^{-\frac12 (\ve x - \veg\mu)^T \vveg\Sigma^{-1} (\ve x - \veg\mu)}.
	\end{eqnarray}
Then it is easy to show
	\begin{eqnarray}
	\rho_\others(x_2,...,x_\dms) 
	= \frac 1 {\sqrt{(2\pi)^{\dms-1} \det\vveg\Sigma_\excl1}}
	e^{-\frac12 (\ve x_\excl1 - \veg\mu_\excl1)^T 
		    \vveg\Sigma_\excl1^{-1} (\ve x_\excl1 - \veg\mu_\excl1)},
	\end{eqnarray}
where $\vveg\Sigma_\excl1$ is the covariance matrix $\vveg\Sigma$
with the first row and first column deleted, and
$\veg\mu_\excl1$ is the vector $\veg\mu$ with the first entry removed. 
For easy correspondence, we will still count the entries as those as 
numbered in $\vveg\Sigma$ and $\veg\mu$. So 
	\begin{eqnarray*}
	&&F_i \Di {\log\rho_\others} \bracket{f_i + \sum_{j=1}^\dms a_{ij} x_j}
		\Di\ \bracket{-\frac12 (\ve x_\excl1 - \veg\mu_\excl1)^T
		\vveg\Sigma_\excl1^{-1} (\ve x_\excl1 - \veg\mu_\excl1) }\\
	&&
	= \parenth{f_i + \sum_{j=1}^\dms a_{ij} x_j} \dot
	  \sum_{j=2}^\dms \parenth{-\frac {\sigma_{ij}'+\sigma_{ji}'} 2 }
			  \cdot (x_j - \mu_j).
	\end{eqnarray*}
Here $\sigma_{ij}'$ is the $(i,j)^{\rm th}$ entry of the matrix
$\vveg\Sigma_\excl1^{-1}$. (Note here the entry indices run 
from 2 through $\dms$, not from 1 through $\dms$!)
As $\vveg\Sigma_\excl1$ is symmetric, so is $\vveg\Sigma_\excl1^{-1}$,
and hence $(\sigma_{ij}'+\sigma_{ji}')/2 = \sigma_{ij}'$.
So
	\begin{eqnarray*}
	&&-E F_i \Di {\log\rho_\others} = 0 - E \sum_{j=1}^\dms a_{ij} x_j 
	      \cdot \sum_{j=2}^\dms (-\sigma_{ij}')\cdot (x_j - \mu_j) \\
	&&= E \sum_{k=1}^\dms a_{ik} (x_k - \mu_k) \cdot
	      \sum_{j=2}^\dms \sigma_{ij}' (x_j - \mu_j)	\\
	&&= \sum_{k=1}^\dms \sum_{j=2}^\dms a_{ik} \sigma_{ij}' 
			E(x_k - \mu_k) (x_j - \mu_j)	\\
	&&= \sum_{k=1}^\dms \sum_{j=2}^\dms a_{ik} \sigma_{ij}'\sigma_kj.
	\end{eqnarray*}
The other term
	\begin{eqnarray*}
	-E \sum_{i=2}^\dms \Di {F_i} = -\sum_{i=2}^\dms a_{ii}.
	\end{eqnarray*}
Eq.~(\ref{eq:T1n_linear}) follows by summing these two terms together.
%\end{proof}
\qed

When $\dms=2$, the above formula can be further simplified. In fact, 
	\begin{eqnarray*}
	T_{1\to2} = a_{21} \sigma_{22}' \cdot \sigma_{12}
		  + a_{22} \sigma_{22}' \cdot \sigma_{22}
		  - a_{22}.
	\end{eqnarray*}
In this case, $\sigma_{22}' = 1/\sigma_{22}$, so
	\begin{eqnarray*}
	T_{1\to2} = a_{21} \frac {\sigma_{12}} {\sigma_{22}},
	\end{eqnarray*}
just as expected (cf.~\cite{Liang2008}).

From above it is easy to see that, 
	\begin{eqnarray}
	T_{1\to\others} \ne \sum_{j=2}^\dms T_{1\to j}.
	\end{eqnarray}
That is to say, the macrostate of a network is not just a simple addition
of the individual states.
The equality can hold only when the $\dms$ components are uncorrelated,
i.e., when $\vveg\Sigma$ is a diagonal matrix, and hence
$\sigma_{ii}'= 1/\sigma_ii$ and $\sigma_{ij}'=0$ for $i\ne j$. 
Indeed, in this case, the $\dms$ components are just independent units;
they do not form a network.

\subsection{The impact of $\ve x_\excl1$ on $x_1$}

We know information flow or causality is asymmetric between two entities;
that is to say, the contribution of $x_1$ to the rest of the network is
generally different from that the other way around. For late reference, 
we here briefly present the
result of the information flow from $\ve x_\excl1$ to $x_1$, though it is
not needed in this study.

From \cite{Liang2008}, 
	\begin{eqnarray}
	\dt {H_1} = - E \bracket{F_1 \DI {\log\rho_1}}
		    - \frac12 E \bracket{g_{11} \DIDI {\log\rho_1} }.
	\end{eqnarray}
Now if we modify the system on the infinitesimal interval $[t+\delt]$ by
freezing $(x_2,x_3,...,x_\dms)$, and follow the above derivation, we
finally arrive at the time rate of change of the marginal entropy of 
$x_1$ with the effect of $(x_2,x_3,...,x_\dms)$ excluded is
	\begin{eqnarray}
	\dt {H_{1,\excl\others}} 
	= E\parenth{\DI {F_1}} - \frac12 E\parenth{g_{11} \DIDI {\log\rho_1}}
	  - \frac12 E\parenth{\frac1{\rho_1} \DIDI {g_{11}\rho_1} }.
	\end{eqnarray}
So the information flow from $\ve x_\excl1$ to $x_1$ is
	\begin{eqnarray}
	T_{\others\to1} 
	&=& \dt {H_1} - \dt {H_{1,\excl\others}}
	= - E\bracket{F_1 \DI {\log\rho_1} + \DI {F_1}}
	  + \frac12 E \bracket{\frac1 {\rho_1} \DIDI {g_{11}\rho_1 }} \cr
	&=& - E\bracket{\frac1 {\rho_1} \DI {F_1\rho_1} }
	  + \frac12 E \bracket{\frac1 {\rho_1} \DIDI {g_{11}\rho_1 }}.
	\end{eqnarray}
A seemingly surprising observation is that this is precisely the same in form 
as that for 2D systems (see \cite{Liang2008}), although here the
dimensionality can be larger than 2. 
This does make sense, as we are splitting the system into two subsystems,
one with $x_1$, another with a collection of $\dms-1$ units.
In the meantime, this generally  differs in form from those individual
information flow formulas for systems with $\dms>2$ (see \cite{Liang2016}).

\section{Maximum likelihood estimation}	\label{sect:mle}
Given a system like (\ref{eq:gov}), we can
rigorously evaluate the information flows among the components.
Now suppose, instead of the system, what we have are just $\dms$ 
time series with $\ntime$ steps, $\ntime\gg\dms$, 
$\{x_1(k)\}, \{x_2(k)\},..., \{x_\dms(k)\}$.
We can estimate the system from the
series, and then apply the information flow formula to fulfill the task.
Assume a linear model as shown above, and assume $m=1$.
following Liang (2014)\cite{Liang2014}, 
the maximum likelihood estimator of $a_{ij}$
is equal to the least-square solution of the following
over-determined problem
	\begin{eqnarray*}
	\left(\begin{array}{ccccc}
	1  &x_1(1)  &x_2(1) &...  &x_n(1) \\
	1  &x_1(2)  &x_2(2) &...  &x_n(2) \\
	1  &x_1(3)  &x_2(3) &...  &x_n(3) \\
	\vdots &\vdots &\vdots \ddots &\vdots\\
	1  &x_1(\ntime)  &x_2(\ntime) &...  &x_n(\ntime) 
	\end{array}\right) 
		\left(\begin{array}{c}
		f_i \\
		a_{i1}\\
		a_{i2}\\
		\vdots\\
		a_{in}
		\end{array}\right)
	=
		\left(\begin{array}{c}
		\dot x_i(1) \\
		\dot x_i(2) \\
		\dot x_i(3) \\
		\vdots\\
		\dot x_i(\ntime) 
		\end{array}\right)
	\end{eqnarray*}
where $\dot x_i(k) = (x_i(k+1)-x_i(k))/\delt$ 
($\delt$ is the time stepsize), 
for $i=1,2,...,\dms$, $k=1,...,\ntime$.  
Use overbar to denote the time mean over the $\ntime$ steps.
The above equation is
	\begin{eqnarray*}
	\left(\begin{array}{ccccc}
	1  &\bar x_1         &\bar x_2        &...  &\bar x_n    \\
	0  &x_1(2)-\bar x_1  &x_2(2)-\bar x_2 &...  &x_n(2)-\bar x_n \\
	0  &x_1(3)-\bar x_1  &x_2(3)-\bar x_2 &...  &x_n(3)-\bar x_n \\
	\vdots &\vdots &\vdots \ddots &\vdots\\
	0  &x_1(\ntime)-\bar x_1  &x_2(\ntime)-\bar x_2 &...&x_n(\ntime)-\bar x_n
	\end{array}\right) 
		\left(\begin{array}{c}
		f_i \\
		a_{i1}\\
		a_{i2}\\
		\vdots\\
		a_{i\dms}
		\end{array}\right)
	=
		\left(\begin{array}{c}
		\bar{\dot x}_i \\
		\dot x_i(2) - \bar{\dot x}_i \\
		\dot x_i(3) - \bar{\dot x}_i \\
		\vdots\\
		\dot x_i(\ntime) - \bar{\dot x}_i
		\end{array}\right)
	\end{eqnarray*}
Denote by $\vve R$ the matrix
	$$\left(\begin{array}{ccccc}
	x_1(2)-\bar x_1  &x_2(2)-\bar x_2 &...  &x_n(2)-\bar x_n \\
	\vdots &\vdots &\vdots \ddots &\vdots\\
	x_1(\ntime)-\bar x_1  &x_2(\ntime)-\bar x_2 &...&x_n(\ntime)-\bar x_n
	\end{array}\right) ,$$
$\ve s$ the vector 
$(x_i(2)-\bar{\dot x}_i, ..., x_i(\ntime)-\bar{\dot x}_i)^T$,
and $\ve a_i$ the row vector $(a_{i1},...,a_{i\dms})^T$.
Then $\vve R \ve a_i = \ve s$. The least square solution of $\ve a_i$,
$\ve {\hat a}_i$, solves 
	\begin{eqnarray*}
	\vve R^T \vve R \ve {\hat a}_i = \vve R^T \ve s.
	\end{eqnarray*}
Note $\vve R^T \vve R$ is $\ntime\vve C$, where $\vve C$ is the covariance
matrix. So
	\begin{eqnarray}
		\left(\begin{array}{c}
		\hat a_{i1}\\
		\hat a_{i2}\\
		\vdots\\
		\hat a_{i\dms}
		\end{array}\right)
	= \vve C^{-1} 
		\left(\begin{array}{c}
		 c_{1,di}\\
		 c_{2,di}\\
		\vdots\\
		 c_{\dms,di}
		\end{array}\right)
	\end{eqnarray}
where $c_{j,di}$ is the covariance between the series $\{x_j(k)\}$  and 
$\{ (x_i(k+1) - x_i(k))/\delt \}$.

So finally, the mle of $T_{1\to\others}$ is 
	\begin{eqnarray}	\label{eq:T1n_est}
	\hat T_{1\to\others} 
	= \sum_{i=2}^\dms \bracket{
	  \sum_{j=2}^\dms c_{ij}' \parenth{\sum_{k=1}^\dms \hat a_{ik} c_{kj}}
		- \hat a_{ii}
				  },
	\end{eqnarray}
where $c_{ij}'$ is the $(i,j)^{\rm th}$ entry of $\tilde{\vve C}^{-1}$, and
	\begin{eqnarray}
	\tilde {\vve C} =
	\left(\begin{array}{ccccc}
	1  &0                &0               &...  &0           \\
	0  &c_{22}  &c_{23} &...  &c_{2\dms} \\
	0  &c_{23}  &c_{33} &...  &c_{3\dms} \\
	\vdots &\vdots &\vdots &\ddots &\vdots\\
	0  &c_{2\dms} &c_{3\dms} &...&c_{\dms\dms}
	\end{array}\right). 
	\end{eqnarray}
Denoting by $\hat{\vve A}$ the matrix with entries $(\hat a_{ij})$, 
Eq.~(\ref{eq:T1n_est}) can be more succinctly written as:
	\begin{eqnarray}	\label{eq:T1n_est2}
	\hat T_{1\to\others}
	= Tr_\excl1 \bracket{\tilde{\vve C}^{-1} (\hat{\vve A} \vve C)^T}  
	-  Tr_\excl1\bracket{\hat {\vve A}}.
	\end{eqnarray}
Here $Tr_\excl1$ means the trace of a matrix with the first term
removed. That is to say, it is defined such that, for matrix $\vve Q$, 
	$$Tr_\excl1 \vve Q = Tr\vve Q - Q(1,1).$$
Note this is made possible by the form of $\tilde {\vve C}$ (with its
special form in $1^{st}$ row and $1^{st}$ column); otherwise the trace of
the product of two matrices, say, $\vve P_{n\times n} \vve Q_{n\times n}
\equiv \vve R_{n\times n}$, is generally not equal to 
$Tr[\vve P(2:n,2:n) \vve Q(2:n,2:n)] + R(1,1)$.

\section{Application to a network of coupled Stuart-Landau oscillators} \label{sect:landau}
%\subsection{Syncrhonization with R\"ossler system} chaotic oscillators
%
% Following \cite{Palus2018}, we consider the following coupled R\"ossler
% systems
%	\begin{eqnarray*}
%	&&\dt{x_1} = -\omega_1 x_2(t) - x_3(t),	\\
%	&&\dt{x_2} = \omega_1 x_1(t) + a x_2(t),	\\
%	&&\dt{x_3} = b + x_3(t) [x_1(t) - c]	\\
%	\end{eqnarray*}
%%
%	\begin{eqnarray*}
%	&&\dt{y_1} = -\omega_2 y_2(t)-y_3(t)  + \epsilon_y[x_1(t)-y_1(t)],\\
%	&&\dt{y_2} = \omega_2 y_1(t) + a y_2(t),	\\
%	&&\dt{y_3} = b + y_3(t) [y_1(t) - c]		\\
%	\end{eqnarray*}
%%
%	\begin{eqnarray*}
%	&&\dt{z_1} = -\omega_3 z_2(t) - z_3(t) + \epsilon_z[x_1(t)-z_1(t)],\\
%	&&\dt{z_2} = \omega_3 z_1(t) + a z_2(t),	\\
%	&&\dt{z_3} = b + z_3(t) [z_1(t) - c]	\\
%	\end{eqnarray*}
%%
%where $a=0.2$, $b=0.2$, $c=5.7$, as chosen by R\"ossler for a chaotic
%attractor. As Palu$\rm\check s$\cite{Palus2018}, we choose $\alpha_j$ 
%around $1$: $\omega_1=1.02$, $\omega_2=0.98$, $\omega_3=0.96$.
%Clearly, $x$ is the driving, master system, while $y$ and $z$ are slaves.
%$y$ and $z$ are not directly connected. $\epsilon_y$ and $\epsilon_z$ are
%two coupling coefficients; here we set both of them to be 0.15.

%%chaos is needed to evaluate information flow

\begin{figure}[htp]
\begin{center}
  \includegraphics[width=0.65\textwidth]{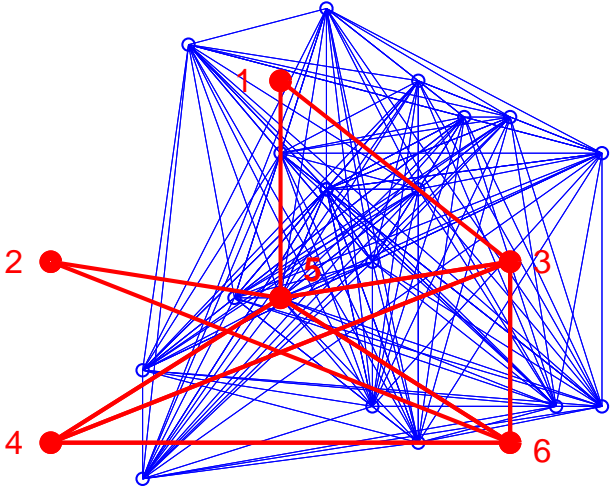}\\
  \caption{A schematic of the network of coupled oscillators.
	For the sake of clarity, 
	in this study only the 6-node (red) subnetwork is considered. }\label{fig:network}
  \end{center}
\end{figure}

In this section we put (\ref{eq:T1n_est}) to application to
a network with $N$ nodes, each made of a Stuart-Landau 
oscillator\cite{Strogatz2001}.
This has been used to model many biological networks for phenomena such
as circadian rhythms, synchronized neuronal firing, and 
spatiotemporal activity in the heart and the brain 
(see \cite{Tanaka2012} for more examples).
For the purpose of demonstration, here a small number $N=6$ is chosen.
Let the complex state variable of the $j^{\rm th}$ oscillator be $z_j$.
It is defined as (see, e.g., \cite{Daido2004}\cite{Tanaka2012}) 
	\begin{eqnarray}	\label{eq:Stuart-Landau}
	\dt {z_j} = \parenth{\alpha_j + i \Omega_j - |z_j|^2} z_j
	+ \frac K N \sum_{k=1}^N  \Adj_{jk} (z_k - z_j)
	+ \nu\dot w_j,
	\qquad j=1,...,N,
	\end{eqnarray}
where $i=\sqrt{-1}$, $\Omega_j$ are the frequencies, 
$\alpha_j$ are control parameters,
and $(\Adj)$ is the adjacency matrix.
Here the coupling coefficient $K$ is chosen to be 1. 
The notation generally follows that in \cite{Tanaka2012}; the difference
lies in an $\Omega$ varying oscillator by oscillator, and an additional
stochastic term $\nu\dot w_j$, where $w_j$ is a standard Wiener process,
and $\nu$ the stochastic perturbation amplitude.
We add some weak stochasticity for convenience (see below).
If $K=0$ and $\nu=0$, the oscillators are Stuart-Landau oscillators; 
a positive $\alpha_j$ yields an oscillating state, whereas a negative
$\alpha_j$ disable the oscillator.  % (see Fig.~\ref{fig:series_single}).
In this study, $K=1$, $\Omega_j = j/2$, $j=1,...,N$, are fixed throughout.
$\alpha_j$ may be $1$ or $-3$, depending on whether $z_j$ is 
activated or switched off. The adjacency matrix is chosen such that 
   $\Adj_{2k}=\Adj_{k2}=0$, $k=1,3,4$;  $\Adj_{1k}=\Adj_{k1}=0$, $k=4,6$,
   and for all other $(j,k)$, $\Adj_{jk}=1$.
The resulting network is sketched in Fig.~\ref{fig:network}.
Obviously, $z_5$ is a highly connected node, or hub; second to it is $z_6$. 
$z_1$ and $z_2$ are two sparsely connected nodes.

\begin{figure}[htp]
\begin{center}
  \includegraphics[width=1\textwidth]{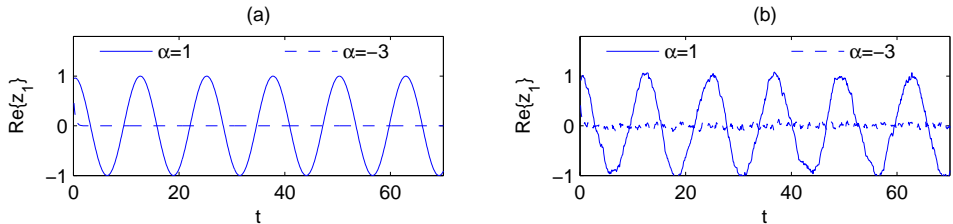}\\
  \caption{Time series of a single oscillator $z_1$ without coupling 
	($K=0$). (a) No noise; (b) weak stochasticity applied ($\nu=0.1$).
	   Only the real parts are drawn.
	}\label{fig:series_single}
  \end{center}
\end{figure}

Equation~(\ref{eq:Stuart-Landau}) is discretized and solved using the 
second order Runge-Kutta scheme.
The system is initialized with random values, integrated forward with a
time stepsize of $\delt=0.1$. Without coupling, the individual oscillators
operate on their own, each exhibiting a periodic series with a distinct
frequency. Shown in Fig.~\ref{fig:series_single}a are the
active (solid) and inactive (dashed) modes for $z_1$ when $\nu=0$. 
Fig.~\ref{fig:series_single}b displays the corresponding cases when
$\nu=0.1$. We need this slightly perturbed system because, as seen in
Fig.~\ref{fig:series_single}a, the trajectories are too regular (periodic),
only leaving on the Poincar'e plane one point. 
In other words, they 
contain no information, making the information flow problem singular. 
Recently it is found this is actually an extreme case\cite{Liang2020}, 
and hence can be handled by perturbing the system slightly with weak
stochasticity. (In real systems, noises are ubiquitous.)
The Fig.~\ref{fig:series_single}b approximate well 
its deterministic case, Fig.~\ref{fig:series_single}a, except
for some weak ripples superimposed on the curves.
So it is reasonable to believe that the addition of the
weak perturbation can be used to compute the information flow
for the original system.

\begin{figure}[htp]
\begin{center}
  \includegraphics[width=1\textwidth]{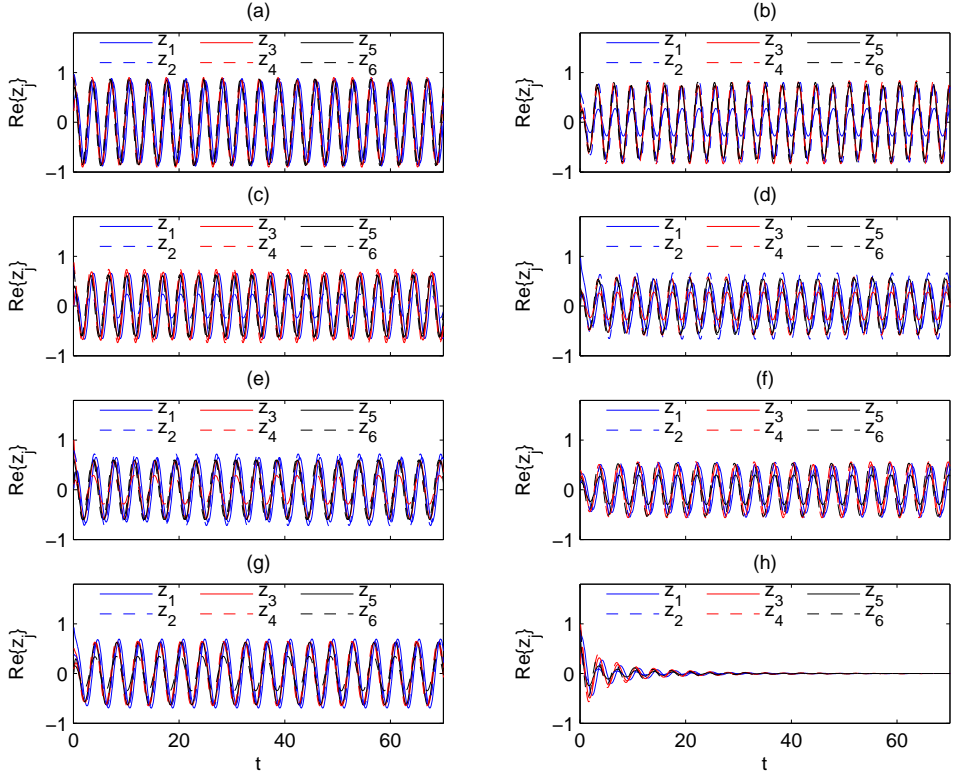}\\
  \caption{Time series of the six coupled oscillators $z_j$
	  (only the real parts of $z_j$ are drawn). 
	(a) All oscillators are active;
	(b) $z_1$ inactive (all others are active; same below);
	(c) $z_2$ inactive;
	(d) $z_3$ inactive;
	(e) $z_4$ inactive;
	(f) $z_5$ inactive;
	(g) $z_6$ inactive;
	(h) both $z_2$ and $z_5$ are inactive;
	}\label{fig:series6}
  \end{center}
\end{figure}

Figure~\ref{fig:series6} shows the time series of the six coupled oscillators. 
In (a), all of them are on. 
As seen, though the frequencies $\Omega_j$ differ, 
the six oscillators work together to produce completely synchronized
oscillations (see \cite{Sun2019} for optimum synchronizations).
To assess the importance of a node, a usual practice is to delete it
from the network and observe the response. 
In Figs.~\ref{fig:series6}b-g, shown are the respective responses 
when $z_1$-$z_6$ are turned off respectively.
Obviously, with only one node failure the network is still alive. But
one can see that the impact of $z_5$ is significantly larger than others, 
while that from $z_1$ is by far the least. In Fig.~\ref{fig:series6}h, 
when $z_5$ and $z_2$ are disabled, then the entire network gradually dies,
though in this case $\alpha_1, \alpha_3, \alpha_4, \alpha_6$ are still positive.
% The network blackout

As mentioned in the introduction,
the above assessment by preferential removal of designated node(s) 
may not be feasible for many networks in nature, neuronal networks in
particular. Now
use formula (\ref{eq:T1n_est}) to estimate the information flow from
the individual oscillators to the network. To begin, note that each $z_j$ 
actually has two components; so they should be taken as two time series.
That is to say, the dynamical system has a dimensionality of $2\times N$.
The remaining computation is straightforward. We generate series with 5000
steps, with the first 100 steps discarded (to ensure stationarity). 
The computed results are 
(units in nats per unit time; 
values may differ slightly due to the random initialization):
	\begin{center}
	\begin{tabular}{cccccc}
	\hline
	\hline
	 $\hat T_{1\to{network}}$ & $\hat T_{2\to{network}}$ 
	  & $\hat T_{3\to{network}}$\ & $\hat T_{4\to{network}}$\ 
	  & $\hat T_{5\to{network}}$\ & $\hat T_{6\to{network}}$ \\
	0.66 &1.30  & 2.11  & 2.50 & 3.07  &3.03 \\
	\hline
	\end{tabular}
	\end{center}
By comparison $z_5$ and $z_6$ are most important; second to them are
$z_3$ and $z_4$. $z_1$ and $z_2$ are least important. 
The result is just as that as illustrated in Fig.~\ref{fig:series6}.
From out common intuition, this makes sense, too. 
As we can check from Fig.~\ref{fig:network}, 
$z_5$ and $z_6$ are the hubs, whereas $z_1$ and $z_2$ are sparsely
connected. 

However, if there exist directed links and/or localized weights (e.g.,
\cite{Sun2019}) in the network, hubs need not always be the most crucial units.
To see this, let $\Adj_{52} = 10$, $\Adj_{62} = 5$. The computed result is
tabulated as follows:
	\begin{center}
	\begin{tabular}{cccccc}
	\hline
	\hline
	 $\hat T_{1\to{network}}$ & $\hat T_{2\to{network}}$ 
	& $\hat T_{3\to{network}}$\ & $\hat T_{4\to{network}}$\ 
	& $\hat T_{5\to{network}}$\ & $\hat T_{6\to{network}}$ \\
	0.50 & 4.00  & 1.92  & 2.10 &2.55 &0.77	\\
	\hline
	\end{tabular}
	\end{center}
So now the most important node is $z_2$, though it is sparsely connected!
And, the impact from $z_6$ has been greatly reduced.

\begin{figure}[htp]
\begin{center}
  \includegraphics[width=1\textwidth]{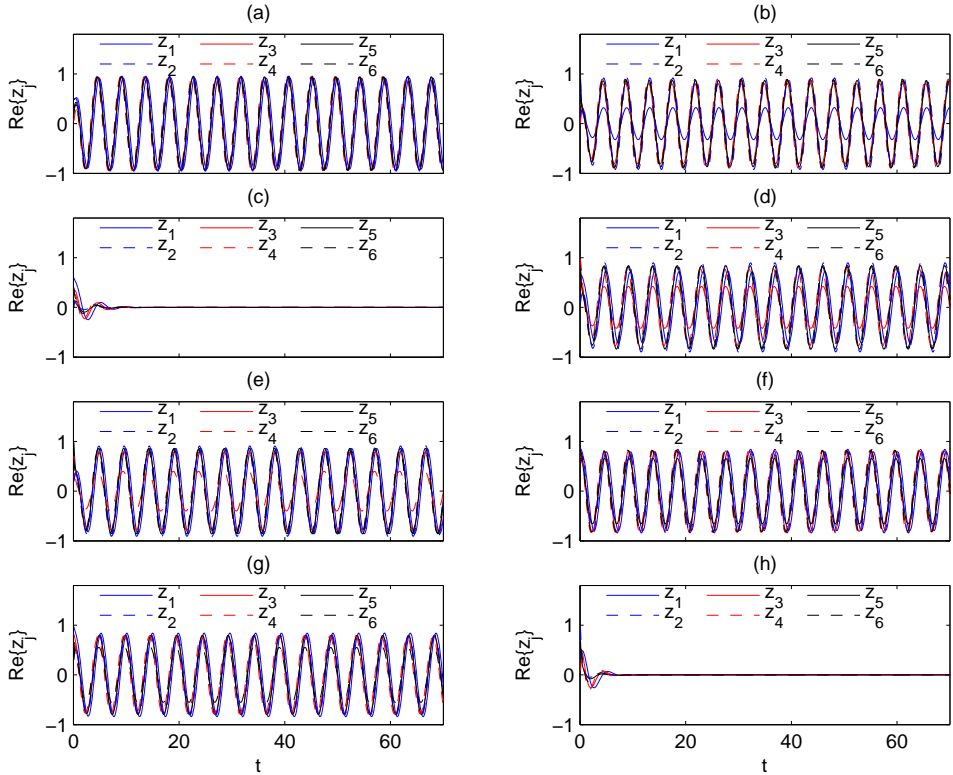}\\
  \caption{
	As Fig.~\ref{fig:series6}, but with weighted and directed links.
	}\label{fig:series6_weighted}
  \end{center}
\end{figure}

To see whether this is indeed the case, we do the node removal experiments
again. Indeed, if $z_2$ is deteriorated or suppressed, the whole network 
becomes silent, as shown in Fig.~\ref{fig:series6_weighted}c.
The result is hence validated.

\section{Summary}	\label{sect:summary}

A quantitative evaluation of the contribution of individual units in 
producing the collective behavior of a complex network is important in that
is allows us to gain an understanding of which units determine the
vulnerability of the network.
In this study, we show that a a natural measure is the information 
flow from the unit in concern to the entire network. 
A formula is derived, and its maximum likelihood estimator provided.
The results are summarized henceforth for easy reference.

For a network modeled with an $\dms$-dimensional continuous-time
dynamical system
	\begin{eqnarray*}
	\dt {\ve x} = \ve F (\ve x, t) + \vve B(\ve x, t) \dot{\ve w},
	\end{eqnarray*}
the information flow from node~$x_1$ to the network $x_2,x_3,...,x_\dms$ 
is
	\begin{eqnarray*}
	T_{1\to\others}
	= -E\bracket{\sum_{i=2}^\dms \frac1 {\rho_\others}
				     \Di{F_i\rho_\others}  }
	  + \frac12 E\bracket{
		\sum_{i=2}^\dms \sum_{j=2}^\dms \frac 1 {\rho_\others}
			\DiDj {g_{ij} \rho_\others }
			     }.
	\end{eqnarray*}
When only time series are available, under the assumption of linearity, 
the maximum likelihood estimator of $T_{1\to\others}$ is
	\begin{eqnarray*}	
	\hat T_{1\to\others}
	= Tr_\excl1 \bracket{\tilde{\vve C}^{-1} (\hat{\vve A} \vve C)^T}  
	-  Tr_\excl1\bracket{\hat {\vve A}}.
	\end{eqnarray*}
In the equation, $Tr_\excl1$ means the trace of a matrix with the first term removed,
$\vve C=(c_{ij})$ is the covariance matrix, $\tilde{\vve C}$ is equal to
$\vve C$ except 
$\tilde c_{1,1}=1$, $\tilde c_{j,1}= \tilde c_{1,j}=0, j=2,3,...,\dms$.
$\hat {\vve A} = (\hat a_{ij})$ has entries 
	\begin{eqnarray*}
		\left(\begin{array}{c}
		\hat a_{i1}\\
		\hat a_{i2}\\
		\vdots\\
		\hat a_{i\dms}
		\end{array}\right)
	= \vve C^{-1} 
		\left(\begin{array}{c}
		 c_{1,di}\\
		 c_{2,di}\\
		\vdots\\
		 c_{\dms,di}
		\end{array}\right), \qquad i=1,2,...,\dms,
	\end{eqnarray*}
where $c_{j,di}$ is the covariance between the series $\{x_j(k)\}$  and 
$\{ (x_i(k+1) - x_i(k))/\delt \}$.
Observe that this ``cumulative information flow" is not equal to the 
sum of the information flows to other individual units, reflecting the 
collective phenomenon that a group is not the addition of the individual 
members. 

The above formula has been put to application to a network consisting of 
Stuart-Landau oscillators. It is shown that the node with largest 
information flow is indeed most crucial for the network. Its deterioration
or suppression will cause the whole network to cease to function. 
An observation is: depending on the topology, such a node may not be a hub;
on the contrary, it could be some sparsely connected, low-degree node.
This study is expected to be useful in identifying clues to 
the mystery why initially small shocks at some nodes may trigger a massive, 
global shutdown of the entire network.

\section*{Acknowledgments} 
%The author thanks the NUIST High Performance Computing Center for
%providing the computing resource.
This study is supported by the National Science Foundation of China (Grant
\# 41975064) and the 2015 Jiangsu Program for Innovation Research and
Entrepreneurship Groups.

\medskip
% The data information below will be filled by AIMS editorial staff
Received xxxx 20xx; revised xxxx 20xx.
\medskip

\end{document}